# One-Pot Multi-component Synthesis of 1,4-Dihydropyridine Derivatives in Biocompatible Deep Eutectic Solvents

**Suhas Pednekar*, Rahul Bhalerao, Nitin Ghadge**

*Organic Chemistry Research Laboratory, Department of Chemistry,

Ramnarain Ruia College, Matunga, Mumbai-400 019, India.

(Tel: + 91-22-24143098 Fax: + 91-22-24143119 email: nitin.ghadge3@gmail.com)

## Abstract

An efficient protocol for the synthesis of differently substituted 1, 4-dihydropyridines in deep eutectic solvents under solvent-free conditions is reported herewith. Excellent yields of the resultant products have been obtained. Recyclability studies have also been performed for deep eutectic solvents with very little loss in activity up to five recycles.

## 1. Introduction

1,4-Dihydropyridines (1,4-DHPs) are an important class of bioactive molecules, well known for their role as calcium channel modulators and used extensively for the treatment of hypertension.[1-3] The derivatives of 1,4-DHPs have shown a variety of biological activities such as vasodilator, bronchodilator, antitumor, hepatoprotective and geroprotective activity.[4,5] Commercial drugs such as Nifedipine which are a prototype of the 1,4-DHP structure has been used extensively in both antianginal and antihypertensive treatment.[6]

One-pot multi-component reactions are of increasing academic and ecological interest due to the possibility of achieving high synthetic efficiency and reaction design.[7] The classical method for the synthesis of 1,4-DHPs involves the one-pot condensation of three or more components under relatively harsh conditions using ammonia as the nitrogen source.[8] Drastic reduction in reaction times and improved reactivity of the synthetic procedures have been brought about by the application of microwave.[9] Although, impressive levels of activity could be achieved, large scale synthesis is a problem. Similarly, Lewis acid catalysts such as $Yb(OTf)_3$ have been shown to catalyze the reactions at ambient temperature.[10]

Recently, Zhao *et.al.* reported the synthesis of 1,4-DHP derivatives in ionic liquids.[11] Ionic liquids have attracted lot of attention during the past decade in the context of green chemistry due to their low vapour pressure, high thermal stability and their solvation properties which make them such unique solvents in synthesis.[12] However, serious limitations have emerged for the industrial scale application of ionic liquids such as high cost,[13] environmental toxicity[14] and the demand for high purity which could otherwise lead to change in the physical properties of the ionic liquids.[15,16] The development of alternative solvents from components that are inexpensive, non-toxic towards the environment and are biodegradable is therefore highly desirable to overcome these drawbacks.

One such alternative approach could be the use of deep eutectic mixtures (DES) which were first synthesized by Abbott *et.al.*[17] and since then found wide scope in an array of biological applications due to their biodegradable properties. Most common of them are the biological transformations such as hydrolase catalyzed biotransformation[18] and extraction of glycerol from biodiesel into a eutectic-based solvent.[19] Although attractive, deep eutectic mixtures have seldom been used for organic transformations.[20] We therefore thought it would be interesting to try the application of DES for one-pot synthesis of 1, 4-dihydropyridines and it worked very efficiently. A mixture of aldehyde **2a-n**, dimedone **3**, ethyl acetoacetate **4** and ammonium acetate **5** in DES **1a** on heating resulted in the formation of 1, 4-dihydropyridines **6a-n.**

## 2. Experimental

### **2.1** *Materials*

All common reagents and solvents were used as obtained from commercial supplies without further purification.

### **2.2** *Apparatus*

All melting points were determined in open capillaries using Gallen Kamp melting point apparatus and are uncorrected. All PMR spectra were recorded in $CDCl_3$ (chemical shift in delta) using Bruker AV 200 spectrophotometer (300 MHz).

### **2.3** *Representative procedure for the preparation of Deep Eutectic Solvent 1a.*

Choline chloride **A** (2.8 g, 20 mmol) and urea **B** (**a**) (2.4 g, 40 mmol) were placed in a round bottomed flask and heated to 60°C, until liquid began to form. After 10 min, a homogenous colourless liquid was formed which was used directly for the reactions.

### **2.4** *Representative procedure for the synthesis of 1, 4-DHP Derivatives 6a-n.*

**2a-n** (1.0 mmol), **3** (0.140 g, 1.0 mmol), **4** (0.130 g, 1.0 mmol), **5** (0.115 g, 1.5 mmol) and DES (Deep Eutectic Solvent) **1a** (2.0 mL) were successively charged into a schlenk tube equipped with a magnetic stirrer. Then the reaction was stirred at 60 °C for 20 minutes and a solid product gradually formed. After the completion of reaction as indicated by TLC, water was added, the reaction mass was stirred and the mixture was filtered. The crude product obtained was purified by recrystallization using absolute alcohol. Further water was removed under vacuum at 80 $^{0}$C leaving behind deep eutectic solvent which was used up to 5 recycles. A detailed account of the recycle studies is mentioned in results and discussion.

**2.5 Spectroscopic data of selected compounds**

**Ethyl-1,4,5,6,7,8-hexahydro-2,7,7-trimethyl-5-oxo-4-phenylquinoline-3-carboxylate (6a).** m.p: 203–204$^{0}$C. IR (KBr in cm$^{-1}$) 3233, 3210, 3080, 1696, 1602, 1059, 692, $^{1}$H NMR ($CDCl_3$, 300 MHz): δ 0.94 (s, 3H), 1.07 (s, 3H), 1.21 (t, *J* = 7.1 Hz, 3H), 2.13–2.29 (m, 4H), 2.35 (s, 3H), 4.06 (q, *J* = 7.1 Hz, 2H), 5.07 (s, 1H), 6.25 (s,1H), 7.08–7.13 (m, 1H), 7.18–7.23 (m, 2H), 7.28–7.33 (m, 2H). ESI-MS (*m/z*) 340 $(M+H)^{+}$. Analysis calculated for $C_{21}H_{25}NO_3$: C, 74.31; H, 7.42; N, 4.13. Found: C, 74.27; H, 7.39; N, 4.08.

**Ethyl-1,4,5,6,7,8-hexahydro-2,7,7-trimethyl-5-oxo-4-p-tolylquinoline-3-carboxylate (6b).** m.p: 260–262$^{0}$C. IR (KBr in cm$^{-1}$) 3287, 3031, 2958, 1644, 1596, 1480, 1334, 1213, 757, $^{1}$H NMR ($CDCl_3$, 300 MHz): δ 0.94 (s, 3H), 1.08 (s, 3H), 1.21 (t, *J* = 7.1 Hz, 3H), 2.10–2.24 (m, 4H), 2.26 (s, 3H), 2.37 (s, 3H), 4.06 (q, *J* = 7.1 Hz, 2H), 5.03 (s, 1H), 5.96 (s, 1H), 7.02 (d, *J* = 8.0 Hz, 2H), 7.19 (d, *J* = 8.0 Hz, 2H). ESI-MS (*m/z*) 354 $(M+H)^{+}$. Analysis calculated for $C_{22}H_{27}NO_3$: C, 74.76; H, 7.70; N, 3.96. Found: C, 74.74; H, 7.65; N, 3.91.

**Ethyl-1,4,5,6,7,8-hexahydro-4-(4-methoxyphenyl)-2,7,7-trimethyl-5-oxoquinoline-3-carboxylate (6c).** m.p: 256-257$^{0}$C. IR (KBr in cm$^{-1}$) 3281, 3199, 3080, 1708, 1607, 1224, 837, $^{1}$H NMR ($CDCl_3$ 300 MHz,): δ 0.94 (s, 3H), 1.07 (s, 3H), 1.21 (t, *J* = 7.2 Hz, 3H), 2.09-2.20 (m, 4H), 2.36 (s, 3H), 3.73 (s, 3H), 4.04 (q, *J* = 7.2 Hz, 2H), 4.99 (s, 1H), 6.06 (s, 1H), 6.65 (d, *J* = 7.3 Hz, 2H), 7.24 (d, *J* = 7.3 Hz, 2H ). ESI-MS (*m/z*) 370 $(M+H)^{+}$. Analysis calculated for $C_{22}H_{27}NO_4$: C, 71.54; H, 7.31; N, 3.79; Found: C, 71.59; H, 7.35; N, 3.84.

**Ethyl-1,4,5,6,7,8-hexahydro-4-(4-hydroxyphenyl)-2,7,7-trimethyl-5-oxoquinoline-3-carboxylate (6d).** m.p: 232–234$^{0}$C. IR (KBr in cm$^{-1}$) 3285, 1690, 1617, 1229, $^{1}$H NMR ($CDCl_3$, 300 MHz): δ 0.94 (s, 3H), 1.08 (s, 3H), 1.19 (t, *J* = 7.2 Hz, 3H), 2.09–2.19 (m, 4H), 2.34 (s, 3H), 4.06 (q, *J* = 7.6 Hz, 2H), 4.72 (s, 1H), 4.98 (s,1H), 5.61 (s, 1H), 6.65 (d, *J* = 8.9

Hz, 2H), 7.17 (d, *J* = 8.4 Hz, 2H). ESI-MS (*m/z*) 356 $(M+H)^+$. Analysis calculated for $C_{21}H_{25}NO_4$: C, 70.98; H, 7.04; N, 3.94; Found: C, 70.96; H, 7.02; N, 3.93.

**Ethyl-1,4,5,6,7,8-hexahydro-2,7,7-trimethyl-4-(4-nitrophenyl)-5-oxoquinoline-3-carboxylate (6e).** m.p: 241-243$^0$C. IR (KBr in cm$^{-1}$) 3290, 3150, 1711, 1657, $^1$H-NMR ($CDCl_3$ 300 MHz,): δ 0.9 (s, 3H), 1.08 (s, 3H), 1.82 (t, *J* = 7.1 Hz, 3H), 2.16-2.36 (m, 4H), 2.40 (s, 3H), 4.01 (q, *J* = 7.1 Hz, 2H), 5.15 (s, 1H), 6.31 (s, 1H), 7.51 (d, *J* = 8.2 Hz, 2H), 8.09 (d, *J* = 8.3 Hz, 2H). ESI-MS (*m/z*) 385 $(M+H)^+$. Analysis calculated for $C_{21}H_{25}N_2O_5$ C, 65.45; H, 6.49; N, 7.27. Found: C, 65.43; H, 6.49; N, 7.28.

**Ethyl-4-(2-chlorophenyl)-1,4,5,6,7,8-hexahydro-2,7,7-trimethyl-5-oxoquinoline-3-carboxylate (6f).** m.p: 208-210$^0$C. IR (KBr in cm$^{-1}$) 3062, 2955, 1720, 1640, 1610, 1468, 1385, 1228, 1020, 745, $^1$H NMR ($CDCl_3$, 300 MHz): δ 0.95 (s, 3H), 1.07 (s, 3H), 1.17 (t, *J* = 7.1 Hz, 3H), 2.15–2.23 (m, 4H), 2.30 (s, 3H), 4.04 (q, *J* = 7.1 Hz, 2H), 5.38 (s, 1H), 6.11 (s, 1H), 6,98 (m, 1H), 7.15–7.19 (m, 1H), 7.24–7.26 (m, 1H), 7.41-7.44 (m, 1H). ESI-MS (*m/z*) 374 $(M+H)^+$. Analysis calculated for $C_{21}H_{24}ClNO_3$: C, 67.43; H, 6.47; N, 3.75. Found: C, 67.44; H, 6.48; N, 3.72.

**Ethyl-4-(4-(dimethylamino)phenyl)-1,4,5,6,7,8-hexahydro-2,7,7-trimethyl-5-oxoquinoline-3-carboxylate (6i).** m.p: 229-231$^0$C. IR (KBr in cm$^{-1}$) 3287, 3018, 2964, 1602, 1510, 1377, 1207, 751, $^1$H NMR ($CDCl_3$, 300 MHz): δ 0.96 (s, 3H), 1.07 (s, 3H), 1.22 (t, *J* = 7.1 Hz, 3H), 2.18–2.27 (m, 4H), 2.35 (s, 3H), 2.86 (s, 6H), 4.05 (q, *J* = 7.1 Hz, 2H), 4.95 (s, 1H), 5.88 (s, 1H), 6.60 (d, *J* = 8.6 Hz, 2H), 7.16 (d, *J* = 8.6 Hz, 2H). ESI-MS (*m/z*) 383 $(M+H)^+$. Analysis calculated for $C_{23}H_{30}N_2O_3$: C, 72.22; H, 7.91; N, 7.32. Found: C, 72.18; H, 7.85; N, 7.25.

**Ethyl-4-(4-chlorophenyl)-1,4,5,6,7,8-hexahydro-2,7,7-trimethyl-5-oxoquinoline-3-carboxylate (6j).** m.p: 245–246$^0$C. IR (KBr in cm$^{-1}$) 3275, 3200, 3075, 2965, 1705, 1650, 1605, $^1$H NMR ($CDCl_3$, 300 MHz): δ 0.93 (s, 3H), 1.08 (s, 3H), 1.19 (t, *J* = 7.1 Hz, 3H),

2.15–2.23 (m, 4H), 2.38 (s, 3H), 4.04 (q, *J* = 7.1 Hz, 2H), 5.02 (s, 1H), 5.90 (s, 1H), 7.15–7.19 (m, 2H), 7.24–7.26 (m, 2H). ESI-MS (*m/z*) 374 $(M+H)^+$. Analysis calculated for $C_{21}H_{24}ClNO_3$: C, 67.46; H, 6.47; N, 3.75. Found: C, 67.43; H, 6.49; N, 3.73.

**Ethyl-1,4,5,6,7,8-hexahydro-2,7,7-trimethyl-4-(3-nitrophenyl)-5-oxoquinoline-3-carboxylate (6k).** m.p: 177–178$^0$C. IR (KBr in cm$^{-1}$) 3303, 2954, 1683, 1610, 1167, 759, $^1$H NMR ($CDCl_3$, 300 MHz): δ 0.89 (s, 3H), 1.05 (s, 3H), 1.22 (t, *J* = 7.1 Hz, 3H), 2.12–2.23 (m, 4H), 2.38 (s, 3H) 4.03 (q, *J* = 7.1 Hz, 2H), 5.16 (s, 1H), 6.85 (s, 1H), 7.36 (t, *J* = 7.9 Hz, 1H), 7.71 (d, *J* = 7.9 Hz, 1H), 7.97 (m, 1H), 7.99 (m, 1H). ESI-MS (*m/z*) 385 $(M+H)^+$. Analysis calculated for $C_{21}H_{25}N_2O_5$: C, 65.45; H, 6.49; N, 7.27. Found: C, 65.44; H, 6.48; N, 7.26.

**Ethyl-1,4,5,6,7,8-hexahydro-2,4,7,7-tetramethyl-5-oxoquinoline-3-carboxylate (6l).** m.p: 201–203$^0$C. IR (KBr in cm$^{-1}$) 3287, 2947, 1685, 1147, 751, $^1$H NMR ($CDCl_3$, 300 MHz): δ 0.98 (d, 3H), 1.09 (s, 6H), 1.26-1.31 (t, 3H), 2.11–2.32 (m, 7H), 3.91-3.93 (q, 1H), 4.17-4.18(m, 2H), 5.6 (s, 1H). ESI-MS (*m/z*) 278 $(M+H)^+$. Analysis calculated for $C_{16}H_{23}NO_3$: C, 69.31; H, 8.30; N, 5.05. Found: C, 69.29; H, 8.29; N, 5.06.

**Ethyl-4-ethyl-1,4,5,6,7,8-hexahydro-2,7,7-trimethyl-5-oxoquinoline-3-carboxylate (6m).** m.p: 143-145$^0$C. IR (KBr in cm$^{-1}$) 3288, 2954, 1686, 1167, 759, $^1$H NMR ($CDCl_3$, 300 MHz) : δ 0.74 (t, *J* = 7.2 Hz 3H), 1.10 (s, 6H), 1.26-1.30 (m, 5H), 2.12–2.34 (m, 7H), 4.02 (t, *J* = 6.2 Hz, 1H), 4.16 (m, 2H), 5.58 (s, 1H). ESI-MS (*m/z*) 292 $(M+H)^+$. Analysis calculated for $C_{17}H_{25}NO_3$: C, 70.10; H, 8.59; N, 4.89. Found: C, 70.12; H, 8.57; N, 4.87.

**Ethyl-4-butyl-1,4,5,6,7,8-hexahydro-2,7,7-trimethyl-5-oxoquinoline-3-carboxylate (6n).** m.p: 165-167$^0$C. IR (KBr in cm$^{-1}$) 3290, 3090, 1690, 1197, 751, $^1$H NMR ($CDCl_3$, 300 MHz): δ 0.82 (t, *J* = 7.1 Hz 3H), 1.09 (s, 6H), 1.15-1.41 (m, 9H), 2.19–2.30 (m, 7H), 4.01 (t, *J* = 6.1 Hz, 1H), 4.17 (m, 2H), 5.77 (s, 1H). ESI-MS (*m/z*) 320 $(M+H)^+$. Analysis calculated for $C_{19}H_{29}NO_3$: C, 71.47; H, 9.09; N, 4.38. Found: C, 71.45; H, 9.07; N, 4.40.

## 3. Results and Discussions

We first set out to prepare series of deep eutectic solvents as mixtures of **A** with amides **Ba-c** (ratio 1:2) and **A** with carboxylic acids **Bd-f** (ratio 1:1) (Scheme-1,Table 1).[17]

We first tested DES **1a-f** as solvents for the synthesis of 1, 4-DHPs in a one-pot multi-component reaction between **2a-n**, **3, 4** and **5**. A direct comparison of DES as solvent, with water and under solventless conditions was also carried out. Initial experiments suggested that the reactions initiated in DES **1a-f** gave better yields of the desired products at an optimum temperature of 60$^{o}$C as compared to water as solvent and under solventless conditions. Interestingly, **1a** proved superior even to others, enabling the synthesis of the 1, 4-DHPs in excellent yield (95.0%, Entry 1, Table 2). Polar and non-polar organic solvents (Entry 11-13, Table 2) when employed as solvents under similar reaction conditions showed poor conversion to the 1, 4-DHPs. These results highlight the potential of deep eutectic solvents **1a-f** for promoting synthetically important transformations in comparison to the conventional solvent systems. Easy work-up procedures could be employed to obtain the product for deep eutectic solvents as they could be easily washed away with water suggesting their biocompatible nature.

Encouraged by the above results we probed the applicability of DES **1a** as the optimum solvent for the synthesis of 1, 4-DHPs in **Scheme-2**. A variety of substituted aldehydes underwent smooth conversion to the desired products. Presence of electron-withdrawing or electron-donating substituents had no influence on the outcome of the reaction with both giving excellent yields. Unsubstituted benzaldehyde (Entry 1, Table 3) gave the best results followed by electron-withdrawing –$NO_2$ (Entry 5, Table 3) and –Cl (Entry 6, Table 3) groups. The electron-donating –OMe (Entry 3, Table 3) or –Me (Entry 2, Table 3) weren't far behind. Highly substituted aldehydes such as vanillin (Entry 7, Table 3) and 5-bromo vanillin (Entry 8, Table 3) also gave well to excellent yields of the 1, 4-DHP products. It was further observed that even aliphatic aldehydes gave products in good yields (Entry 12-13, Table 3). The generality of the protocol is thus exhibited allowing efficient synthesis of differently substituted 1, 4-DHPs.

### 3.1 Recycle study of Deep eutectic Solvent

The ability to recover and recycle the solvent used in reaction is an important aspect to green chemistry. Therefore in present study we also evaluated the recyclability and reusability of this deep eutectic solvent (Table 4).

## 4. Conclusion

In summary a very simple approach towards one-pot multi-component synthesis of 1, 4-Dihydropyridine Derivatives was developed using Biocompatible Deep Eutectic solvents. The deep eutectic solvents gave excellent recyclability without loss of activity up to 5 cycles hence providing a greener alternative. A milder and highly efficient protocol was developed which can easily replace existing methods.

References

1. Bossert F, Meyer H and Wehinger E 1981 *Angew. Chem. Int. Ed. Engl.* **20** 762
2. Mannhold R, Jablonka B, Voigdt W, Schoenafinger K and Schrava K 1992 *Eur. J. Med. Chem.* **27** 229
3. Reid G L, Meredith P A and Pasanisi F 1985 *J. Cardiovasc. Pharmacol.* **7** S18
4. Shan R, Velazquez C and Knaus E 2004 *J. Med. Chem.* **47** 254
5. Kawase M, Shah A, Gaveriya H, Motohashi N, Sakagami H, Varga A and Molnar J. 2002 *Bioorg. Med. Chem.* **10** 1051
6. (a) Janis R A and Triggle D J 1983 *J. Med. Chem.* **26** 775; (b) Loev B, Goodman M M, Snader K M, Tedeschi R and Macko E 1974 *J. Med. Chem.* **17** 956

7. (a) Tietze L F 1996 *Chem. Rev.* **96** 115; (b) Tietze L F, Brasche G and Gericke K M, 2006 *Domino Reactions in Organic Synthesis*; Wiley-VCH: Weinheim,
8. Love B and Sander K M 1965 *J. Org. Chem.* **30** 1914
9. Tu S-J, Yu C-X, Liu X –H, Yao C –S, Liu F and Gao Y 2002 *Chin. J. Struct. Chem.* **21** 99
10. Wang L-M, Sheng J, Zhang L, Han J –W, Fan Z –Y, Tian H and Qian C –T 2005 *Tetrahedron* **61** 1539
11. Ji S J, Jiang Z Q, Lu J and Loh T P 2004 *Synlett*. 831
12. (a) Welton T 1999 *Chem. Rev*. **99** 2071; (b) Ranke J, Stolte S, Stormann R, Arning J and Jastorff B 2007 *Chem. Rev.* **107** 2183; (c) Jastorff B, Stormann R, Ranke J, Molter K, Stock F, Oberheitmann B, Hoffmann W, Hoffmann J, Nuhter M, Ondruschka B and Filser J 2003 *Green Chem.* **5** 136; (d) Stolte S, Arning J, Bottin-Weber U, Matzke M, Stock F, Thiele K, Uerdingen M, Welz-Biermann U, Jastorff B and Ranke J 2006 *Green Chem*. **8** 621–629; e) Anastas P, Wasserscheid P and Stark A 2010 *Green Solvents: Ionic Liquids*, Eds: Wiley-VCH Weinheim
13. Park S and Kazlauskas R J 2003 *Curr. Opin. Struct. Biol.* **14** 432
14. Gericke M, Fardim P and Heinze P 2012 *Molecules* **17** 7458
15. (a) Aparicio S, Atilhan M, Karadas F 2010 *Ind. Eng. Chem. Res.* **49** 9580; (b) Seddon K R, Stark A and Torres M J 2000 *Pure Appl. Chem.* **72** 2275; (c) Ab Rani M A, Brant A, Crowhurst L, Dolan A, Lui M, Hassan N H, Hallett J P, Hunt P A Niedermeyer H and Perez-Arlandis J M 2011 *Phys. Chem. Chem. Phys.***13** 16831; (d) Gallo V, Mastrorilli P, Nobile C F, Romanazzi G and Suranna G P 2002 *J. Chem. Soc. Dalton* **23** 4339; (e) Lee S, Ha S, Lee S and Koo Y M 2006 *Biotechnol. Lett.* **28** 1335
16. Shamsuri A A and Abdullah D K 2010 *MAKARA, SAINS.* **14** 101

17. (a) Capper G, Davies D, Munro H, Rasheed R and Tambyrajah V 2001 *Chem. Commun.* 2010; (b) Abbott A, Capper G, Davies D, Rasheed R and Tambyrajah V 2003 *Chem. Commun*. 70 ;(c) Abbott A, Capper G, Davies D and Rasheed R. 2004 *J. Am. Chem. Soc* **126** 9142
18. Gorke J, Romas J and Kazlauskas H 2008 *Chem. Commun*. 1235
19. Abbott A, Cullis P, Gibson M, Harris R and Raven E 2007 *Green Chem.* **9** 868
20. Jain N, Kumar A and Chauhan S 2005 *Tetrahedron* **61** 1015